\documentclass[pra, twocolumn,showpacs,amsmath,amssymb]{revtex4}

\usepackage{graphicx}%Include figure files
\usepackage{dcolumn}%Align table columns on decimal point
\usepackage{bm}% bold math
\begin{document}

\title{Blackbody-radiation shift in a  $^{88}$Sr$^+$ ion optical frequency standard}

\author{Dansha Jiang}
\affiliation{Department of Physics and Astronomy, University of
Delaware, Newark, DE 19716-2570, USA}
\author {Bindiya Arora}
\affiliation{Department of Physics and Astronomy, University of
Delaware, Newark, DE 19716-2570, USA}
\author{M. S. Safronova}
 \homepage{http://www.udel.edu/~msafrono}
 \email{msafrono@udel.edu}
\affiliation{Department of Physics and Astronomy, University of
Delaware, Newark, DE 19716-2570, USA}
\author{Charles W. Clark}
\affiliation{
Joint Quantum Institute, National Institute of Standards and Technology
and the University of Maryland, Gaithersburg, Maryland 20899-8410, USA
}

\begin{abstract}
The blackbody radiation (BBR) shift of the $5s - 4d_{5/2}$ clock transition in $^{88}$Sr$^+$
 is calculated to be $0.250(9)$~Hz at room temperature, $T=300$~K,
  using the relativistic all-order method where all single and double
 excitations of the Dirac-Fock wave function are included to all orders of perturbation theory.
 The BBR shift is a major component in the uncertainty budget of the
 optical frequency standard based on the $^{88}$Sr$^+$ trapped ion.
The scalar polarizabilities of the $5s$ and $4d_{5/2}$ levels,
as well as the tensor polarizability of the $4d_{5/2}$ level,
are presented together with the evaluation of their uncertainties.
The lifetimes of the $4d_{3/2}$, $4d_{5/2}$, $5p_{1/2}$, and $ 5p_{3/2}$ states
are calculated and compared with experimental values.
\end{abstract}

 \maketitle

\section{Introduction}
The current definition of a second in the International System of
Units (SI)  is based on the microwave transition between the two
hyperfine levels of the ground state of $^{133}$Cs.
The present relative standard uncertainty of Cs microwave frequency
standard is around $4\times10^{-16}$ \cite{NIST-Csclock}. Since the frequencies of
feasible optical clock transitions are five orders of magnitude
greater than the standard microwave transitions,  optical frequency
standards may achieve even smaller  relative uncertainties.
Significant recent progress in optical spectroscopy and
 measurement techniques has led to the achievement of relative standard uncertainties in
  optical frequency standards that are comparable to the Cs microwave benchmark.
  
In 2006, the International Committee for Weights and Measures (CIPM)
 recommended  that the following transitions frequencies shall be used 
 as secondary representations of the second \cite{CIPM1}: 
 ground-state hyperfine microwave transition in $^{87}$Rb \cite{Rb1,Rb2}, 
 $5s\ ^2S_{1/2} -4d\ ^2D_{5/2}$ optical transition of the  $^{88}$Sr$^+$ ion~\cite{Madej,
Margolis}, $5d^{10} 6s \ ^2S_{1/2} (F=0) - 5d^{9} 6s^2 \ ^2D_{5/2} (F=2) $
optical transition in $^{199}$Hg$^+$ ion~\cite{Hg1,Hg2},
$6s \ ^2S_{1/2} (F=0) - 5d \ ^2D_{5/2} (F=2)$ optical transition in $^{171}$Yb$^+$
  ion~\cite{Yb1,Yb2} and  $5s^2 \ ^1S_{0} - 5s5p \ ^3P_{0}$ transition in 
      $^{87}$Sr neutral atom~\cite{Sr1, Sr2,Sr3}.
       With extremely low systematic perturbations
and better stability and accuracy, such optical frequency standards
can reach a systematic fractional uncertainty of the order of
$10^{-18}$~\cite{uncertainty, Margolis}. More precise frequency standards
will open ways to improve global positioning systems and tracking of
deep-space probes, and perform more accurate measurements of the
fundamental  constants and testing of physics postulates.

In this paper, we treat  one of the optical transitions recommended
as secondary standard: the $5s\ ^2S_{1/2} -4d\ ^2D_{5/2}$
 electric-quadrupole  transition of $^{88}$Sr$^+$ at
 $445$ THz ($674$~nm).
The reported frequency measurements of this transition in a single
trapped $^{88}$Sr$^+$ ion
 have achieved spectral resolution of better than
 1.5~Hz \cite{Madej, Madej99, Margolis, Margolis03}.
Methods based on this transition have the  potential to  reduce
relative systematic uncertainty to the level of $10^{-17}$
 or below~\cite{Margolis}.

The accuracy of optical frequency standards is limited by the
frequency shift in the clock transition caused by the interactions
of the ion with external fields. The linear Zeeman shift in the
$^{88}$Sr$^+$ system
 can be eliminated by use of the line center of  symmetric Zeeman
 states; the second-order Zeeman shift is around 1 mHz and is negligible at
current level of precision~\cite{Madej}. The second order Doppler
shifts due to micromotion of the trapped ion are estimated to be
less than 0.01 Hz~\cite{Margolis}. The major contributions to the
systematic frequency shifts come from Stark shifts with the
 blackbody radiation (BBR) Stark shift being one of the most important contributions at room temperature.
Experimental measurements of the BBR radiation shifts are difficult.
Here, we present
 theoretical calculations that result in estimates of this shift that reduce the previous
 uncertainty \cite{Madej,Margolis} by a factor of 10.

In this paper, we present a relativistic all-order calculation of
the static polarizabilities of the  $5s_{1/2}$ and $4d_{5/2}$ states
of $^{88}$Sr$^+$. The relativistic all-order method used here is one
of the most accurate methods used for the calculation of atomic
properties of monovalent systems. Readers are referred to
Ref.~\cite{review07} and references therein for a review of this
method and its applications. We use these polarizability values to
evaluate the BBR shift of the clock transition at room temperature.
The dynamic correction to the electric-dipole contribution and
multipolar corrections due to M1 and E2 transitions are
incorporated. The uncertainty of the final BBR shift is estimated to
be 3.6\%. Lifetimes of the low-lying excited $4d_{3/2}$, $4d_{5/2}$,
$5p_{1/2}$, and $5p_{3/2}$ states are also calculated and compared
with experiments.

\section{Method}
The electrical field $E$ radiated by a blackbody at temperature $T$,
as given by Planck's law,
\begin{equation}
E^2(\omega) d \omega=\frac{8 \alpha^3}{\pi} \frac{\omega^3
d\omega}{\mathrm{exp}(\omega/k_B T)-1},
\end{equation}
induces a nonresonant perturbation of the optical transition at room
temperature~\cite{BBR_Farley}. Assuming that the system evolves
adiabatically, the frequency shift of an atomic state due to such an
electrical field can be related to the static electric-dipole
polarizability $\alpha_0$ by
 (see Ref.~\cite{BBR_Porsev})
\begin{equation}
 \Delta_v = -\frac{1}{2}(831.9~{\textrm V/m})^2
\left( \frac{T(K)}{300} \right)^4 \alpha_0(1+\eta).
\end{equation}
Here,  $\eta$ is a small dynamic correction due to the frequency
distribution. Only the electric-dipole transition part of the
contribution is considered in the formula because the contributions
from M1 and E2 transitions are suppressed by a factor of $\alpha^2$
\cite{BBR_Porsev}. We estimate these multipolar corrections together
with the dynamic correction $\eta$ in Section~\ref{bbr} of this
work. The overall BBR shift of the clock transition frequency is
then calculated as the difference between the BBR shifts of the
individual levels involved in the transition:
\begin{eqnarray} \label{eq:BBR}
 \Delta_{\mathrm{BBR}}(5s\rightarrow 4d_{5/2})& =& -\frac{1}{2}
[\alpha_0(4d_{5/2})-\alpha_0(5s_{1/2})]  \nonumber \\
&\times & (831.9{\rm V/m})^2 \left( \frac{T(K)}{300} \right)^4.
\end{eqnarray}
Therefore, the evaluation of the BBR shift requires accurate
calculation of static scalar polarizabilities of the $^{88}$Sr$^+$
in the $5s_{1/2}$ ground and $4d_{5/2}$ excited states. The static
tensor polarizability of the $4d_{5/2}$ state is also calculated in
the present work though the effect of the tensor part of
polarizability is averaged out due to the isotropic nature of the
electric field radiated by the blackbody.
\begin{table}
\caption{\label{tab-5s}
Contributions to the $5s\ ^2S_{1/2}$ scalar ($\alpha_0$) static polarizabilities
in $^{88}$Sr$^+$ and their uncertainties in units of $a^3_0$.
The values of corresponding $E1$ matrix elements are given in $ea_0$.}
\begin{ruledtabular}
\begin{tabular}{lcc}
Contribution  & $\langle k\Vert D \Vert 5s_{1/2} \rangle $ & $\alpha_0$\\
\hline
$5s_{1/2}-5p_{1/2}$&  3.078 &  29.23 \\
$5s_{1/2}-6p_{1/2}$&  0.025 &  0.001  \\
$5s_{1/2}-7p_{1/2}$&  0.063 &  0.004  \\
$5s_{1/2}-8p_{1/2}$&  0.054 &  0.003   \\
$5s_{1/2}-5p_{3/2}$&  4.351 &  56.48 \\
$5s_{1/2}-6p_{3/2}$&  0.034 & 0.002 \\
$5s_{1/2}-7p_{3/2}$&  0.053 & 0.003 \\
$5s_{1/2}-8p_{3/2}$&  0.054 & 0.003 \\
$\alpha_{\mathrm{core}}$ &&5.81 \\
$\alpha_{vc}$ &&-0.26 \\
$\alpha_{\mathrm{tail}}$ && 0.02\\
$\alpha_{\mathrm{total}}$ && 91.30\\
\end{tabular}
\end{ruledtabular}
\end{table}

The calculation of the  scalar polarizability of a monovalent
atom can be separated into three parts: the contribution of the
electrons in the ionic core, $\alpha_{\mathrm{core}}$; a small term,
$\alpha_{vc}$, that changes the core polarizability due to the
presence of the valence electron; and the dominant  contribution,
$\alpha_{v}$, from the valence electron. The ionic core
polarizability used here was calculated using the random-phase
approximation (RPA) \cite{acore}. We calculate the $\alpha_{vc}$
contribution in the RPA approximation as well for consistency with
the ionic core value. The valence scalar $\alpha_0$ and tensor $\alpha_2$
 polarizabilities of an atom in a state $v$ can be expressed as the sum 
 over all intermediate states $k$ allowed by
the electric-dipole selection rules:
\begin{eqnarray}
\alpha_0 & = &\frac{2}{3(2j_v+1)}\sum_k
\frac{\langle k\Vert D \Vert v\rangle^2}{E_k-E_v}, \\
\alpha_2 &=& -4C \sum_k (-1)^{j_v+j_k+1}
\left\lbrace
\begin{array}{ccc}
j_v & 1 & j_k \\
1 & j_v & 2
\end{array}
 \right\rbrace
\frac{\langle k \Vert D \Vert v \rangle^2}{E_k-E_v},  \nonumber \\
C &=& \left(  \frac{5j_v(2j_v-1)}{6(j_v+2)(2j_v+1)(2j_v+3)}  \right)  ^{1/2} ,
\end{eqnarray}
where $\langle k \Vert D \Vert v \rangle $ are the reduced
electric-dipole (E1) matrix elements and $E_i$ is the energy of the
$i$th state. We also separate the valence polarizability into two
parts, the main term $\alpha_{\mathrm{main}}$ containing the first
few dominant contributions and the remainder
$\alpha_{\mathrm{tail}}$. We use electric-dipole matrix elements
calculated using the relativistic single-double (SD) all-order
method (see \cite{review07} for detail) and experimental energies
from Ref.~\cite{Moore} for the calculations of the main term. Triple
excitations are included partially where needed; the resulting
values are referred to as SDpT (single, double, partial triple)
data.

 The
tail contribution for the $5s$ state is negligible and is estimated
in the lowest-order Dirac-Fock (DF)
 approximation. Significantly larger tail contribution to the $4d_{5/2}$ polarizability
  is evaluated in both Dirac-Fock (DF) and RPA approximation
and scaled to account for other missing correlation corrections.

\begin{table}
\caption{\label{tab-4d-3} Contributions to the $4d_{5/2}$ scalar
($\alpha_0$) and tensor ($\alpha_2$) static polarizabilities in
$^{87}$Sr$^+$ and their uncertainties in units of $a^3_0$. The
absolute values of corresponding $E1$ reduced matrix elements are
given in $ea_0$.}
\begin{ruledtabular}
\begin{tabular}{lccc}
Contribution  & $\langle k\Vert D \Vert 4d_{5/2} \rangle $ & $\alpha_0$ & $\alpha_2$\\
\hline
$4d_{5/2}-5p_{3/2}$& 4.187&44.16(29)&-44.16(29) \\
$4d_{5/2}-6p_{3/2}$&0.142& 0.012(2)&-0.012(2)\\
$4d_{5/2}-7p_{3/2}$&0.078& 0.003&-0.003\\
$4d_{5/2}-8p_{3/2}$&0.053& 0.001&-0.001\\
$4d_{5/2}-4f_{5/2}$& 0.789&0.329(4)&0.376(4)\\
$4d_{5/2}-5f_{5/2}$& 0.442&0.085(2)&0.97(2)\\
$4d_{5/2}-6f_{5/2}$& 0.297&0.035&0.040\\
$4d_{5/2}-7f_{5/2}$& 0.219&0.018&0.021\\
$4d_{5/2}-8f_{5/2}$&0.157&0.009&0.010\\
$4d_{5/2}-9f_{5/2}$&0.138&0.007&0.008\\
$4d_{5/2}-10f_{5/2}$&0.115&0.005&0.0050\\
$4d_{5/2}-11f_{5/2}$&0.098&0.003&0.004\\
$4d_{5/2}-12f_{5/2}$& 0.085&0.003&0.003\\
$4d_{5/2}-4f_{7/2}$& 3.528&6.576(70)&-2.348(25)\\
$4d_{5/2}-5f_{7/2}$& 1.979&1.699(30)&-0.607(11)\\
$4d_{5/2}-6f_{7/2}$& 1.329&0.698(11)&-0.249(4)\\
$4d_{5/2}-7f_{7/2}$& 0.979&0.360(5)&-0.128(2)\\
$4d_{5/2}-8f_{7/2}$&0.764&0.212(4)&-0.076(2)\\
$4d_{5/2}-9f_{7/2}$&0.619&0.136(2)&-0.049(1)\\
$4d_{5/2}-10f_{7/2}$&0.517&0.093(1)&-0.033\\
$4d_{5/2}-11f_{7/2}$&0.440&0.067(1)&-0.024\\
$4d_{5/2}-12f_{7/2}$&0.381&0.050(1)&-0.018\\
$\alpha_{\mathrm{core}}$ && 5.81(29) \\
$\alpha_{\mathrm{vc}}$ &&-0.40(10)&\\
$\alpha_{\mathrm{tail}}$ &&2.06(20)&-0.59(7)\\
$\alpha_{\mathrm{total}}$ &&62.0(5)&-47.7(3) \\
\end{tabular}
\end{ruledtabular}
\end{table}

\begin{table*}
\caption{\label{tab-matrix}
Reduced electric-dipole transition matrix elements calculated using different approximations:
Dirac-Fock (DF), single-double all-order method (SD), and single-double all-order method
including partial triple-excitation contributions (SDpT); the label ``sc'' indicates the corresponding scaled values.
 All values are given in atomic units.}
\begin{ruledtabular}
\begin{tabular}{lcccccc}
Transition  & DF & SD & SDpT & SD$_{\mathrm{sc}}$ & SDpT$_{\mathrm{sc}}$ & Final\\
\hline
$4d_{5/2} - 5p_{3/2}$& 5.002& 4.150&4.198&  4.187& 4.173&4.187(14)\\
$4d_{5/2} - 4f_{5/2}$&0.964 & 0.779&0.790&   0.789&0.785& 0.789(4)\\
$4d_{5/2} - 4f_{7/2}$& 4.313& 3.486&3.536&   3.528&3.509&3.528(19)\\
$4d_{3/2} - 5p_{1/2}$&3.729 & 3.083&	3.119	&3.112	&3.102	&3.112(10)\\
$4d_{3/2} - 5p_{3/2}$&1.657 &1.369	&1.386&	1.383	&1.378&	1.383(5)\\
\end{tabular}
\end{ruledtabular}
\end{table*}

In this work, we use atomic units (a.u.), in which, $e$, $m_e$, $4 \pi \epsilon_0$ and the reduced
Planck constant $\hbar$ have the numerical value $1$. Polarizability in a.u. has the dimension of volume, and its
numerical values presented here are thus expressed in units of $a_0 ^3$, where $a_0 \approx 0.052918$ nm
is the Bohr radius. The atomic units for $\alpha$ can be converted to SI units via $\alpha / h [$Hz$/(V/m)^2]=
2.48832 \times 10 ^{-8} \alpha [a.u.]$, where the conversion coefficient is $4 \pi \epsilon_0 a_0^3 /h$ and Planck
constant $h$ is factored out.

We have used the B-spline method to construct a finite basis set for radial Dirac equations
 as introduced in Ref.~\cite{Bspline}.
70 B-splines of order $k=8$  are constrained to a spherical cavity with $R=220$ a.u. for each angular momentum.
Such a large cavity is chosen to accurately evaluate as many $4d_{5/2}-nf_{7/2}$
transitions as practically possible  to reduce the uncertainty in the remainder.

\section{Polarizabilities}

Table~\ref{tab-5s} shows the contributions of the individual
transitions to the ground state scalar polarizability $\alpha_0$.
The main contributions are listed separately along with the
respective values of the electric-dipole matrix elements. The tail
contributions are grouped together as $\alpha_{\mathrm{tail}}$. For
the main contributions, we use our \emph{ab initio} SD all-order
values of the matrix elements and experimental energies from
Ref.~\cite{Moore}. The  $5s_{1/2}-5p_{1/2}$ and
$5s_{1/2}-5p_{3/2}$ transitions contribute  over 99.9\% to the valence polarizability and
 94\% of the total polarizability value. The
same calculation of these transitions in Rb agrees with
high-precision experiment to 0.26\% \cite{SDpT}. In fact, the SD
values for the primary $ns-np$ transitions in Li, Na, K, Rb, and Cs
agree with various types of high-precision experiments to 0.1\% -
0.4\% \cite{SDpT}. There is no reason to expect reduced accuracy in
the case of Sr$^+$, and we take the uncertainty of these matrix
element values to be 0.5\%. Unfortunately, we know of no way to
accurately estimate the missing additional contributions to the
dominant correlation correction to these transitions,
unlike the case of the $4d-5p$ and $5s-4d$ transitions, where
semiempirical scaling makes possible an uncertainty estimate that
does not directly depend upon  comparison with experiment.

 The core contribution taken from Ref.~\cite{acore} is estimated to be
accurate to $5\%$, based on the comparison of the RPA  and
experimental polarizability values for noble gases. The tail
contribution is calculated using the DF approximation and is
negligible in comparison with the total polarizability. The error of
the tail is taken to be 100\%. As a result, all uncertainties
 except the ones associated with the $5s-5p$ matrix elements are negligible.
 The resulting final uncertainty of the $5s$
 polarizability is thus estimated to be 1\%. We note that accurate measurement of either $5s-5p$ oscillator
 strengths or $5p$  lifetimes ($5p-4d$ contributions are small and can be accurately calculated)
  will help to significantly reduce this uncertainty.

Table~\ref{tab-4d-3} shows the contributions from the individual
transitions to the $4d_{5/2}$ polarizability. Three types of
transitions contribute to the $4d_{5/2}$  polarizability:
$4d_{5/2}-np_{3/2}$, $4d_{5/2}-nf_{5/2}$, and $4d_{5/2}-nf_{7/2}$.
The sum over the $4d_{5/2}-np_{3/2}$ transitions converges very
quickly with the $4d_{5/2}-5p_{3/2}$ term being overwhelmingly
dominant. We obtain an accurate value for this matrix element
using a semi-empirical scaling procedure that evaluates some classes
of correlation corrections omitted by the current all-order
calculations. The scaling procedure is described in
Refs.~\cite{Matrix, MSthesis}. Briefly, the single valence excitation
coefficients are multiplied by the ratio of the corresponding
experimental and theoretical correlation energies, and the matrix
element calculation is repeated with the modified excitation
coefficients. The scaling procedure is particularly suitable for
this transition because the matrix element contribution containing
the single valence excitation coefficients is dominant in this case
(but not for the $5s-5p$ matrix elements discussed earlier). We
conduct the scaling starting from both SD and SDpT approximations.
 The scaling factors for the SD and SDpT calculations
are different, and we take scaled SD value as the final result for
the $4d_{5/2}-5p_{3/2}$ matrix  element, based on the comparisons of
similar calculations in alkali-metal atoms with
experiments~\cite{US-1,US-2,US-3,US-4}. The absolute values of the
reduced $4d_{5/2}-5p_{3/2}$ matrix elements calculated in  different
approximations
 are summarized in Table~\ref{tab-matrix}, together with four other transitions that represent similar cases.
The uncertainties are determined as the maximum difference between the scaled SD values
 and the \emph{ab initio} SDpT and scaled SDpT values. A notable feature of this table is close
 agreement of the scaled SD and SDpT results.
The sum of the contributions from the $4d_{5/2}-nf_{5/2}$ and $4d_{5/2}-nf_{7/2}$ transitions converges slowly;
therefore we include as many transitions as realistically possible in the main term calculation.
Scaled values are used for the $4d_{5/2}-4f_{5/2}$ and $4d_{5/2}-4f_{7/2}$ transitions in the polarizability
calculations.

The tail contribution of the $4d_{5/2}-nf_{7/2}$ terms is
particularly large; its DF value (3.5~a.u.) is
  $5\%$ of the total polarizability. Therefore, we carry out several additional calculations to accurately
  evaluate the tail contribution and estimate its uncertainty. Since the largest part of the correlation
  correction for the $4d_{5/2}-nf_{7/2}$ transitions with $n > 9$ terms comes from RPA-like terms, the RPA
  approximation is expected to produce a better result than the DF one. We carried out the RPA calculation of the
  tail and obtained a lower value of 2.9~a.u.   We also calculated the last few main terms using
the DF and RPA approximations and compared the results with our
all-order values. We found  that the DF and RPA approximations
overestimate the polarizability contributions by $35\%$ and $28\%$,
relative to DF  and RPA values. To improve our
accuracy, we scale both DF and RPA results by these respective
amounts to obtain a DF-scaled value of 2.26 a.u. and RPA-scaled
value of 2.06~a.u. We take RPA-scaled value as the final one and the
difference of these two values as its uncertainty.

We also list the contributions from various transitions to the
$4d_{5/2}$ tensor polarizability $\alpha_2$ in Table~\ref{tab-4d-3}.
 The $4d_{5/2}-np_{3/2}$ transition gives the dominant contribution to the tensor polarizability.
The tail contribution is smaller yet significant and is obtained by
the same procedure as the tail of the scalar $4d_{5/2}$
polarizability.

\begin{table}
\caption{Contributions to the lifetimes of the $5p_{1/2}$ and
$5p_{3/2}$ states. The transitions rates $A$ are given in
$10^6~$s$^{-1}$ and the lifetimes $\tau$ are given in ns.
\label{tab-lif} }
\begin{ruledtabular}
\begin{tabular}{lrllrl}
 \multicolumn{2}{c}{$5p_{1/2}$}     &
 \multicolumn{1}{c}{}     &
\multicolumn{2}{c}{$5p_{3/2}$}      \\ \hline
   $A(5p_{1/2} - 5s)$       & 128.04 && $A(5p_{3/2} - 5s)$      & 141.29      \\
   $A(5p_{1/2} - 4d_{3/2})$ & 7.54   && $A(5p_{3/2} - 4d_{3/2})$& 0.96    \\
   $\sum A        $         & 135.58 && $A(5p_{3/2} - 4d_{5/2})$& 8.06 \\
                            &        && $\sum A        $        &150.31 &         \\[0.5pc]
    \multicolumn{2}{c}{$\tau(5p_{1/2})$}     &
 \multicolumn{1}{c}{}     &
\multicolumn{2}{c}{$\tau(5p_{3/2})$}      \\                       
   Present         & 7.376&&     Present   & 6.653      \\
    Expt.  ~\cite{Lifetime_Pinnington}  &7.39(7) &&Expt.  ~\cite{Lifetime_Pinnington}& 6.63(7)\\
    Expt.~\cite{Lifetime_Kuske} &    7.47(7)  && Expt.~\cite{Lifetime_Kuske}& 6.69(7)\\
   \end{tabular}
\end{ruledtabular}
\end{table}

\section{Lifetimes}

The contributions to the lifetimes of the $5p_{1/2}$ and $5p_{3/2}$
states are given in Table~\ref{tab-lif}. Experimental energies from
Ref.~\cite{Moore} are used in the evaluation of the transition
rates. The lifetime is calculated as the inverse of the sum of the
appropriate Einstein A-coefficients, which are proportional to the
square of the dipole matrix elements; experimental energies from
Ref.~\cite{Moore} are used. Our results are in excellent agreement
with experimental lifetimes $\tau(5p_{1/2})=7.39(7)$ ns and
$\tau(5p_{3/2})=6.63(7)$ ns by Pinnington \emph{et al.} measured
using a laser-induced fluorescence
method~\cite{Lifetime_Pinnington}, and $\tau(5p_{1/2})=7.47(7)$ ns
and $\tau(5p_{3/2})=6.69(7)$ ns by Kuske \emph{et.al.} measured
using fast-beam-laser technique~\cite{Lifetime_Kuske}.

\begin{table}
\caption{\label{tab-lifetimes}
Lifetimes of the $4d_{3/2}$ and $4d_{5/2}$ levels (s).
}
\begin{ruledtabular}
\begin{tabular}{llll}
Levels&Experiment&Other calculations& Present\\
\hline
$4d_{3/2}$&0.435(4)~\cite{Lifetime_Biemont} &0.443~\cite{Mitroy}&0.441(3)\\
&0.455(29)~\cite{Lifetime_Biemont}&0.426(8)~\cite{Lifetime_Sahoo}\\
&0.435(4)~\cite{Lifetime_Mannervik}&0.422~\cite{Lifetime_Biemont}\\
&&0.441~\cite{Lifetime_Poirier}\\[0.5pc]
$ 4d_{5/2}$&0.3908(16)~\cite{Lifetime_Letchumanan}&0.404~\cite{Mitroy}&0.394(3)\\
&0.408(22)~\cite{Lifetime_Biemont}&0.357(12)~\cite{Lifetime_Sahoo}\\
&0.372(25)~\cite{Lifetime_Madej} &0.384~\cite{Lifetime_Biemont}\\
& &0.396~\cite{Lifetime_Poirier} \\
\end{tabular}
\end{ruledtabular}
\end{table}

\begin{table*}
\caption{\label{tab-final} Comparison of static scalar
polarizabilities for the $5s_{1/2}$ and $4d_{5/2}$ states and
blackbody radiation shift for the $5s_{1/2}-4d_{5/2}$ transition in
$^{88}$Sr$^+$ ion at $T=300$ K. The polarizability values
 are in $a^3_0$ and BBR shift is in Hz. }
\begin{ruledtabular}
\begin{tabular}{lcccccccc}
&Present work & \multicolumn{2}{c}{Ref.~\cite{Mitroy}}&
 Ref.~\cite{Barklem}&Ref.~\cite{Madej}& Ref.~\cite{Patil}&Ref.~\cite{Margolis}\\
\hline
$\alpha_0( 5s_{1/2})$&91.3(9)&89.88&&93.3&84.6(3.6)&91.47\\
$\alpha_0( 4d_{5/2})$& 62.0(5)&61.77&62.92*&57.0&48(12)\\
BBR shift &0.250(9)&0.242&0.233*&0.31&0.33(0.12)&&0.33(9)\\
\end{tabular}
\end{ruledtabular}
*Results are obtained using experimental energies.
\end{table*}

As a further  test of accuracy of our approach and accuracy of our all-order
 $4d$ wave functions, we carry out the calculation of the
 $4d_{3/2}$ and $4d_{5/2}$ lifetimes that requires evaluation of the electric-quadrupole and
 magnetic-dipole transitions.

 The lifetime of the $4d_{3/2}$ state is calculated to be $0.441(3)$~s, where the main contribution comes from
the $E2(4d_{3/2}-5s_{1/2})=11.13(3)$~a.u. matrix element.
The contribution from $4d_{3/2}-5s_{1/2}$ M1 transition is evaluated to be negligible.
The most recent lifetime measurements of $4d_{3/2}$ states of Sr$^+$ include
 $0.435(4)$~s result obtained by using optical pumping and $0.455(29)$~s value obtained by using laser
probing as reported in the  same work~\cite{Lifetime_Biemont}.
Our result agrees with the experimental values within the uncertainty limits.
The lifetime of the $4d_{5/2}$ state is calculated to be $0.394(3)$~s.
The contribution to the A-coefficients from the $4d_{5/2}-5s_{1/2}$ E2 transition overwhelmingly dominates,
 and the corresponding reduced matrix element is $13.75(4)$~a.u.
The reduced matrix element for the $4d_{5/2}-4d_{3/2}$ E2 transition
is $5.98(2)$~a.u., but its contribution to the lifetime is
negligible due to the small energy interval between these two
states. Our value for the $4d_{5/2}-4d_{3/2}$ M1 reduced matrix
element, $1.55$~a.u.,  is in agreement with the result from
 Ref.~\cite{Lifetime_Sahoo}.
The contributions from M1 transitions only affect the $5$th decimal
of the lifetime result of the $4d_{5/2}$ state, and can be neglected at the present level of accuracy.
The two-photon transitions contribute 0.03\% to the $4d$ lifetimes.  
 Our result is found in good agreement with the lifetime measurements of  $0.3908(16)$~s and $0.408(22)$~s
done by Letchumanan \textit{et al.}~\cite{Lifetime_Letchumanan} and Biemont \textit{et al.}~\cite{Lifetime_Biemont}, respectively.
More experimental and theoretical results for the lifetimes of these two states are given in Table~\ref{tab-lifetimes}.

\section{BBR shift}
\label{bbr} We use our scalar polarizability values to evaluate the
shift in the clock transition in $^{88}$Sr$^+$ due to blackbody
radiation at $T=300$~K to be $0.252(9)$~Hz. The dynamic correction
\cite{BBR_Porsev} is estimated to be  $\eta$=0.0013 and $\eta$=0.0064
for the $5s$ and $4d_{3/2}$ states, respectively. The resulting
correction to the BBR shift is $-0.002$~Hz and our final value for
the BBR shift is $0.250(9)$~Hz. The overall uncertainty in the final
result comes from the uncertainty in the values of the $5s-5p_{3/2}$
and $5p_{3/2} - 4d_{5/2}$ matrix elements,  and $4d_{5/2}-nf_{7/2}$
tail. The first two sources of the uncertainties may be removed if
these values were determined experimentally. The $5s-5p_{3/2}$ and
$5p_{3/2} - 4d_{5/2}$ matrix elements can be obtained
 via either lifetime, ground state polarizability, oscillator strength,  or  light shift ratio
measurements  \cite{new-fortson}, with the first two type of
experiments useful for the first matrix element and second two types
for both matrix elements. We note that ionic core uncertainty is not
included in the BBR uncertainty since the core contribution is the
same for both levels and subtracts out. 
We note that the small term $\alpha_{\mathrm{vc}}$ that changes the core polarizability due to the
presence of the valence electron is different for the $5s$ and $4d_{5/2}$ states and contributes 0.5\% to the BBR
shift.
 We also estimated the M1 and
E2 contributions to the BBR shift  using the approach described in
Ref.~\cite{BBR_Porsev} and found them to be negligible (below 0.01\%).

In Table~\ref{tab-final}, we compare our polarizability and BBR
shift results with other theoretical calculations ~\cite{Mitroy,
Barklem, Patil, Madej, Margolis}. We note that our calculation is
the most complete one at present time. The calculation  of Mitroy
\emph{et al.}~\cite{Mitroy} carried out by diagonalizing a
semi-empirical Hamiltonian in a large-dimension single-electron
basis gives the BBR shift of 0.242 Hz. The use of experimental energies 
changes this result to 0.233~Hz \cite{Mitroy}. The accuracy of the scalar polarizability of the ground
state $\alpha_0=89.88$ was estimated to be  $2-3\%$ in
Ref.~\cite{Mitroy}. The results of Barklem and O'Mara~\cite{Barklem}
are derived from experimental oscillator strengths (note that the
core polarizability of $5.8a_0^3$ was added to the values listed in
Ref.~\cite{Barklem}). The results of Madej \emph{et.al.}~\cite{Madej}
are obtained mainly by summing over the transition rates calculated
in Ref.~\cite{Brage} using the multiconfiguration Hartree-Fock
(MCHF) method. It appears that the core contribution (5.8 a.u.) was omitted in \cite{Madej}. 
Addition of the core contribution to the results of \cite{Madej} leads to $\alpha_0(5s) = 90.4$~a.u. and  $\alpha_0(4d_{5/2}) = 54$~a.u. values.
 Omission of the higher-order transition contribution
 resulted in lower values for the scalar polarizability of the $4d_{5/2}$ state in ~\cite{Madej}.
The scalar polarizability of the $5s_{1/2}$ state calculated by Patil and Tang~\cite{Patil}
is obtained by evaluating the transition matrix elements with simple wave functions based on the
asymptotic behavior and on the binding energies of the valence electron.
Our result is in  good agreement with their calculation.
Another estimation of the BBR shift is given by Margolis \emph{et al.} in Ref.~\cite{Margolis},
but the approach is not stated.

\section{Conclusion}
In summary, we calculated the polarizabilities of the $5s$ and
$4d_{5/2}$ states in $^{88}$Sr$^+$ and the value of BBR shift of the
corresponding clock transition at room temperature.
 The dynamic correction to the electric-dipole
contribution and the multipolar corrections due to M1 and E2
transitions were estimated and found to be small at the present
level of accuracy. Lifetimes of the low-lying excited $4d_{3/2}$,
$4d_{5/2}$, $5p_{1/2}$, and $5p_{3/2}$ states were also calculated
and compared with experiments for further tests of our approach.
 The uncertainty of the final BBR value was estimated. The main contributions
 to the uncertainties were
 analyzed and  possible experiments were suggested to further
 reduce the uncertainties of the BBR shift.

This research was performed under the sponsorship of the US Department of Commerce,
National Institute of Standards and Technology.

%\bibliography{BBR_updated}

\end{document}